\documentclass[sigconf, authorversion, nonacm]{acmart}
\setcopyright{none}

\AtBeginDocument{%
  \providecommand\BibTeX{{%
    \normalfont B\kern-0.5em{\scshape i\kern-0.25em b}\kern-0.8em\TeX}}}

\copyrightyear{2023} 
\acmYear{2023} 
\setcopyright{rightsretained}
\acmConference[SIGGRAPH '23 Conference Proceedings]{Special Interest Group on Computer Graphics and Interactive Techniques Conference Conference Proceedings}{August 6--10, 2023}{Los Angeles, CA, USA}
\acmBooktitle{Special Interest Group on Computer Graphics and Interactive Techniques Conference Conference Proceedings (SIGGRAPH '23 Conference Proceedings), August 6--10, 2023, Los Angeles, CA, USA}
\acmDOI{10.1145/3588432.3591526}
\acmISBN{979-8-4007-0159-7/23/08}

\acmSubmissionID{512}

\usepackage{algorithm}
\usepackage[noend]{algpseudocode}
\usepackage{subcaption}
\usepackage{natbib}
\usepackage{tabularx}
\begin{document}

\title{Constructing Printable Surfaces with View-Dependent Appearance}

\author{Maxine Perroni-Scharf}
\affiliation{%
    \institution{Princeton University}
    \streetaddress{88 College Road West}
    \city{Princeton}
    \country{USA}}
\email{maxi@princeton.edu}

\author{Szymon Rusinkiewicz}
    \affiliation{%
    \institution{Princeton University}
    \city{Princeton}
    \country{USA}}
\email{smr@princeton.edu}

\renewcommand{\shortauthors}{Perroni-Scharf}

\begin{abstract}
We present a method for the digital fabrication of surfaces whose appearance varies based on viewing direction. The surfaces are constructed from a mesh of bars arranged in a self-occluding colored heightfield that creates the desired view-dependent effects. At the heart of our method is a novel and simple differentiable rendering algorithm specifically designed to render colored 3D heightfields and enable efficient calculation of the gradient of appearance with respect to heights and colors. This algorithm forms the basis of a coarse-to-fine ML-based optimization process that adjusts the heights and colors of the strips to minimize the loss between the desired and real surface appearance from each viewpoint, deriving meshes that can then be fabricated using a 3D printer. Using our method, we demonstrate both synthetic and real-world fabricated results with view-dependent appearance. 
\\\\A version of this work will appear with the following citation: 
\\Maxine Perroni-Scharf and Szymon Rusinkiewicz. 2023. Constructing Printable Surfaces with View-Dependent Appearance. In Special Interest Group on Computer Graphics and Interactive Techniques Conference Conference Proceedings (SIGGRAPH ’23 Conference Proceedings), August 6–10, 2023, Los Angeles, CA, USA. ACM, New York, NY, USA, 10 pages. https://doi.org/10.1145/3588432.359152
\\\\\\\\
\end{abstract}

\begin{teaserfigure}
  \vspace*{0.05in}
  \begin{center}
  \includegraphics[scale=0.53]{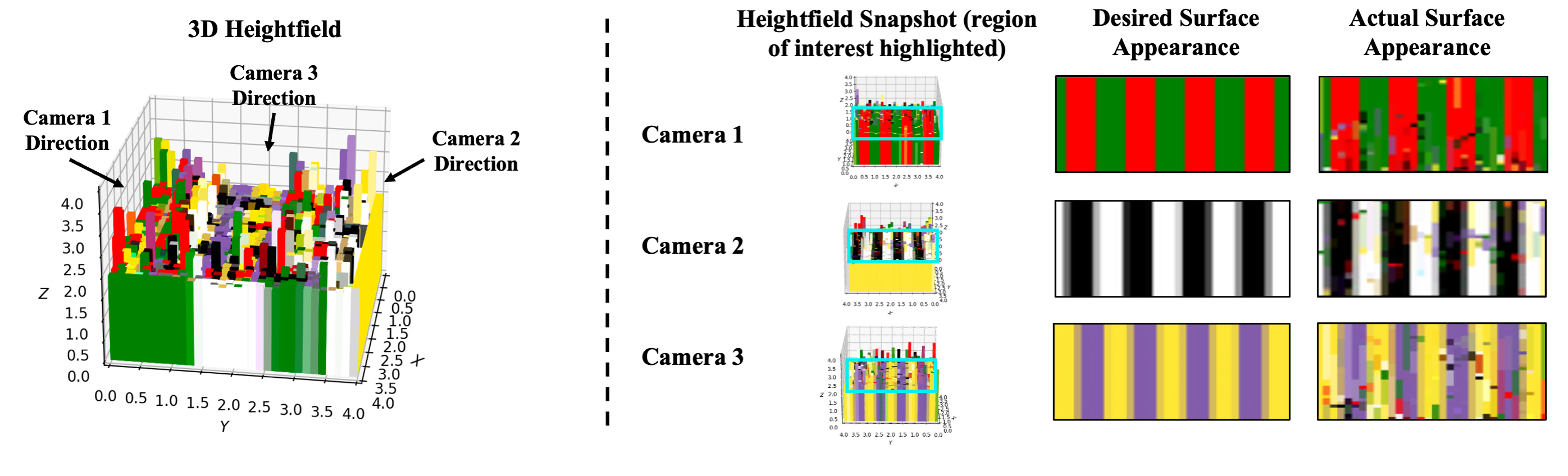}
  \end{center}
  \vspace*{-0.3in}
  \caption{Left: A heightfield in which heights and colors have been optimized by our method in order to produce desired appearances from 3 different directions as indicated. Right: Snapshots of this heightfield from the 3 viewing directions together with the desired and rendered appearances of the regions of interest.}
  \vspace*{0.2in}
  \label{fig:teaser}
\end{teaserfigure}

\renewcommand{\topfraction}{0.9}
\renewcommand{\dbltopfraction}{0.9}
\renewcommand{\bottomfraction}{0.7}
\renewcommand{\floatpagefraction}{0.89}
\renewcommand{\dblfloatpagefraction}{0.89}
\renewcommand{\textfraction}{0}
\setcounter{topnumber}{8}   
\setcounter{dbltopnumber}{5}
\setcounter{bottomnumber}{8}
\setcounter{totalnumber}{12}

\maketitle

\section{Introduction}

Recent advances in 3D printing technology, including the ability to print the entire color spectrum at very high resolutions, have enabled many new applications in digital fabrication. One such application of these capabilities is making directly printable 3D objects whose surface appearance changes dynamically with viewing angle to create multi-image displays. Such tangible displays have applications in various communication tasks and art. For example, one could use such a surface to display co-located AprilTags on one small square \cite{olson2011apriltag}, or to make color changing artwork \cite{weissman20153d, lee2022touch} and interactive objects.

The current state of the art solution for this task, ``Lenticular Objects'' \cite{zeng2021lenticular}, employs a UV printer to print 3D objects covered in tiny lenses. Underneath each lens lies an array of colored dots. The lenses then cause different colors to appear based on viewing direction, allowing multiple images to be displayed on a single object's surface. However, this method requires the use of clear material, and the printed objects appear to have dark a hexagonal overlay around the edges of each lens. These objects also require a polish coating to be painted on them after printing, and the visual effects of the surfaces are very sensitive to the type of coating used. Furthermore, lens-based approaches can only a achieve low resolutions: with lenticular objects, the resolution of the images is limited by the requirement that each lens be large enough to cover a given number of dots (equal to the number of viewing directions). 

Our work aims to achieve higher printable resolutions than prior approaches by using a self-occluding colored heightfield that can be directly fabricated with UV 3D printers and require no additional fabrication steps such as lenses or polish. In the heightfield, certain strips will obstruct each other from different vantage points, creating view-dependent effects such as those shown in Fig. \ref{fig:teaser}. We propose a machine-learning based method for automatically creating these heightfields based on the desired surface appearance. At the heart of our method lies a novel differentiable rendering algorithm tailored specifically to 3D heightfields.

We present a suite of various optimization techniques for self-occluding heightfields. To better search for a global optimum of our system, we use coarse-to-fine stochastic-gradient-descent (SGD) interspersed with simulated annealing. Furthermore, we employ alternating block coordinate descent to improve the accuracy of our results. We regularize our surfaces to make them suitable for 3D printing with extra barrier loss and neighbor loss terms in our objective function. We also conduct a series of experiments to validate the effectiveness of these various techniques and the limitations of our approach in terms of number and range of viewing directions, accuracy and resolution.

In this work we contribute:
\begin{enumerate}
    \item A novel special-purpose differentiable renderer designed for self-occluding heightfields.
    \item A tailored SGD-based optimization algorithm that includes coarse-to-fine surface subdivision, alternating block coordinate descent, steps of simulated annealing, and surface regularization, which is effective for optimizing self-occluding heightfields.
    \item A demonstration of the effectiveness of our algorithm on synthetic and real-world 3D printed results.
\end{enumerate}

\section{Related Work}

\paragraph{View Dependent Appearance and Fabrication}
Previous research employs a variety of methods to achieve surfaces capable of displaying multiple images. \citet{pjanic2015color} propose a metal-printing method that uses superposition of horizontal and vertical colored lines to create surfaces which appear to change color with view direction. Their results, however, have a grainy quality. Also, the approach is specific to metal media and a maximum of two distinct images can be embedded into a single surface.

Self-occlusion is a popular technique for displaying multiple images or changing images on a single surface. \citet{snelgrove2013parallax} employs parallax walls to cause certain colors to appear depending on the direction of incident light on the surface. Unlike our method, the heights of these walls are regular and fixed regardless of the desired input images, making the method constrained to changes in appearance across only one axis. In one of the closest existing works to ours, \citet{sakurai2018fabricating} uses a UV printer to create tiny heightfields that induce self-occlusion by creating several subcells for each desired color at each point in the image and designing walls that block the colors of certain subcells from certain viewing directions. This method, however, compromises image quality, as the black walls and fragments of the other colors are visible from each view angle, creating a grainy texture. This method is also explicitly limited to a low number of views and cannot exploit the benefits of shared pixel color across viewpoints, whereas our method can.

Lenticular lens surfaces can also be used to create surface appearance that changes with view angles. Most recently, \citet{zeng2021lenticular} demonstrate printing lenses directly onto 3D objects to cause changing surface appearance. Due to the nature of the lenses, these surfaces have the appearance of a thin dark hexagonal lattice overlay due to the shadows between the lenticular lenses, whereas our method produces no dark regular artifacts or shadows.

\citet{klehm2014property} and \citet{nindel2021gradient} both propose approaches for respectively optimizing synthetic and 3D printable surfaces by changing the material color and opacity of the surface per voxel to achieve a desired appearance. However, unlike our approach, neither focus on updating surface geometry to create view-dependent appearance.

\paragraph{Self-Occlusion and self-shadows}
The technique of fabricating complex surfaces that have self-occluding or self-shadowing properties has also been previously used in a variety of applications outside of fabricating multi-image displays. \citet{alexa2011images} uses self-shadowing heightfields to produce images that vary based on the direction of incident light. Additionally, \citet{alexa2012irregular} exploits shadowing to dither images by placing irregular pits on surfaces where the deeper the pit, the darker the appearance at that point on the surface. Along these same lines, \citet{peng2019fabricating} fabricates indented 3D surfaces to create QR codes on 3D objects. Our current algorithm does not consider the effects of self-shadowing, but as a future extension of our work it could be augmented to do so using a similar approach to that in \cite{alexa2011images}.

\paragraph{Applications of Digital Fabrication}
The process of applying graphics to the physical world  has been the subject of a longstanding and growing area of research \cite{salisbury1999making, sequin2013making}, with a number of studies focusing in particular on creating surfaces with unique and unusual optical effects. \citet{papas2011goal} uses the technique of indenting surfaces in order to manipulate incident light so that when light shines through the surface, a desired image appears on a projection plane. \citet{regg2010computational} fabricates 3D holograms by indenting parabolic and hyperbolic grooves into specular materials. \citet{pereira2017printing} uses magnetic flakes embedded in resin to print surfaces with anisotropic appearance. \citet{weyrich2009fabricating} optimizes microfacet heightfields to replicate desired reflected highlight shapes, and manufactures these surfaces with a milling machine. 


\paragraph{Differentiable Rendering}
There is a large body of previous work that designs and employs differentiable rendering on meshes for a variety of applications \cite{kato2020differentiable}. To tackle discrete discontinuities that occur from occlusion and object boundaries, one can facilitate gradient computation by approximating the rendering forward pass with a smooth function. For example, \citet{rhodin2015versatile} fades the density of objects at boundaries, and \citet{liu2019soft} proposes the \textit{Soft Rasterizer}, which utilizes spatial blurring and a probabilistic pixel color aggregation method. Recently, \citet{petersen2022gendr} presents a generalized family of differentiable renderers that employ a large variety of sigmoid functions to approximate the Heaviside stepwise function. In our work we also utilize a smooth Heaviside stepwise function approximation for differentiable rendering. However, unlike previous methods, our differentiable renderer is optimized for 3D heightfields. This is a computationally inexpensive approach that allows us to directly compute gradients in terms of the heights and colors of the bars in the heightfield.

\paragraph{3D Reconstruction}
3D reconstruction is a popular computer vision task, with a variety of deep-learning based approaches being used for this purpose \cite{maxim2021survey}. Typically these aim for the appearance of the object to be consistent and coherent across views. In contrast, our work aims for the surface of an object to change its appearance when viewed from different vantage points. For example, NeRF achieves very accurate and consistent results by using a neural radiance field representation for scenes \cite{mildenhall2021nerf} but is not necessarily well-suited for our task, as it does not output an explicit geometry suitable for 3D printing. Algorithms have been proposed to reconstruct 3D meshes from volumetric representations \cite{lorensen1987marching, uy2022point2cyl}, but often suffer from artifacts and introduce an additional phase in the reconstruction process. This is why we instead propose an explicit heightfield optimization approach tailored to creating surfaces with view-dependent appearances.

\begin{figure}
  \centering
  \includegraphics[scale=0.3]{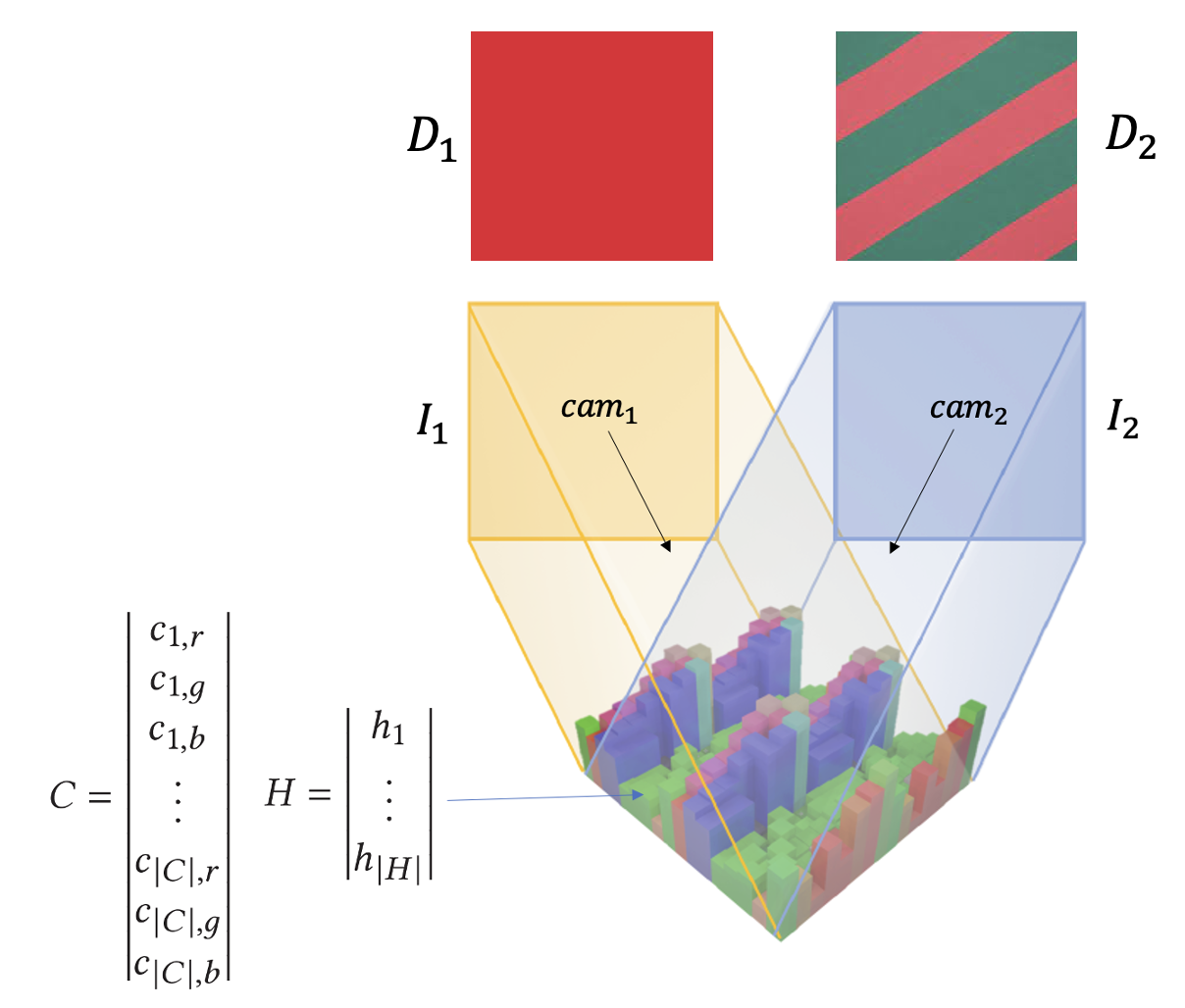}
  \vspace*{-0.2in}
  \caption{Diagram of setup of our system for two arbitrary views. $D_1$ and $D_2$ are desired appearance images. A heightfield with heights $H$ and colors $C$ are viewed by cameras $cam_1$ and $cam_2$ to produce actual appearance images $I_1$ and $I_2$.}
  \vspace*{0.1in}
  \Description{}
  \label{fig:params}
\end{figure}

\section{Heightfield Rendering Algorithm}

We propose a special-purpose differentiable rendering method for 3D heightfields, which we describe below.

\subsection{Heightfield and Camera Setup}

A heightfield is represented by a height matrix $H$ and color matrix $C$. The heightfield is viewed by an series of orthographic cameras $CAM$ pointed at the heightfield, as shown in Fig.~\ref{fig:params}.

$$H = \begin{vmatrix}
h_1\\
\vdots\\
h_{|H|}
\end{vmatrix},\quad
C = \begin{vmatrix}
c_{1,r}\\
c_{1,g}\\
c_{1,b}\\
\vdots\\
c_{|C|,r}\\
c_{|C|,g}\\
c_{|C|,b}\\
\end{vmatrix},\quad
CAM = \begin{vmatrix}
cam_{1,x}\\
cam_{1,y}\\
cam_{1,z}\\
\vdots\\
cam_{|CAM|,x}\\
cam_{|CAM|,y}\\
cam_{|CAM|,z}
\end{vmatrix}.
$$

We use weak perspective projections due to the relatively shallow nature of our heightfields. Each camera $cam_i$ is simply parameterized by a 3D viewing direction vector. For each of these viewing directions, there is a given desired image $D$. Below, we describe the rendering algorithm to find the actual projected image $I$ for each camera.

\begin{figure*}
  \centering
  \vspace*{-0.1in}
  \includegraphics[scale=0.35]{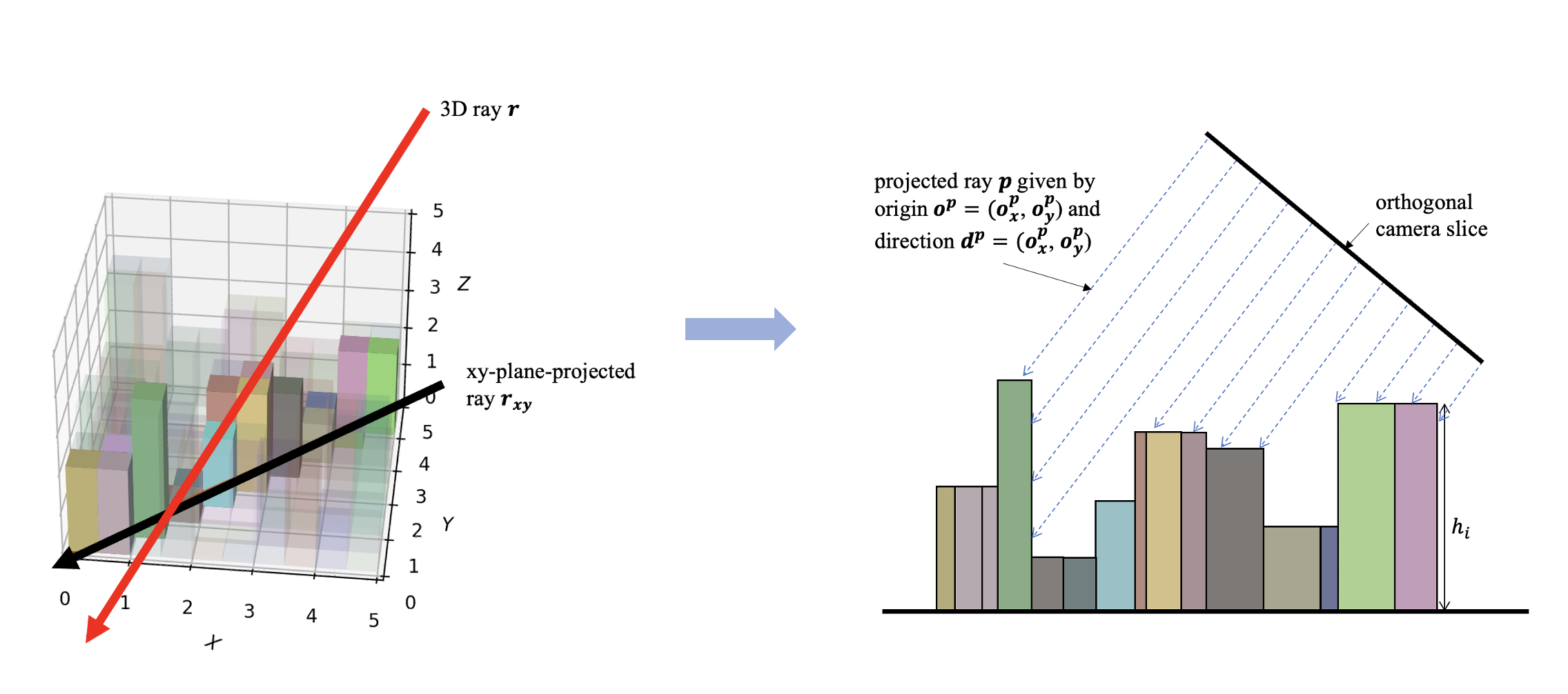}
  \vspace*{-0.3in}
  \caption{Left: a 3D ray $r$ (red) hitting 3D heightfield, with its projected ray (black) and the strips the projected ray intersects highlighted. Right: the corresponding 2D slice of the heightfield with multiple projected camera rays hitting it.}
  \Description{A woman and a girl in white dresses sit in an open car.}
  \label{fig:slicing}
  \vspace*{0.1in}
\end{figure*}

\begin{figure}
  \centering
  \includegraphics[width=0.2\textwidth]{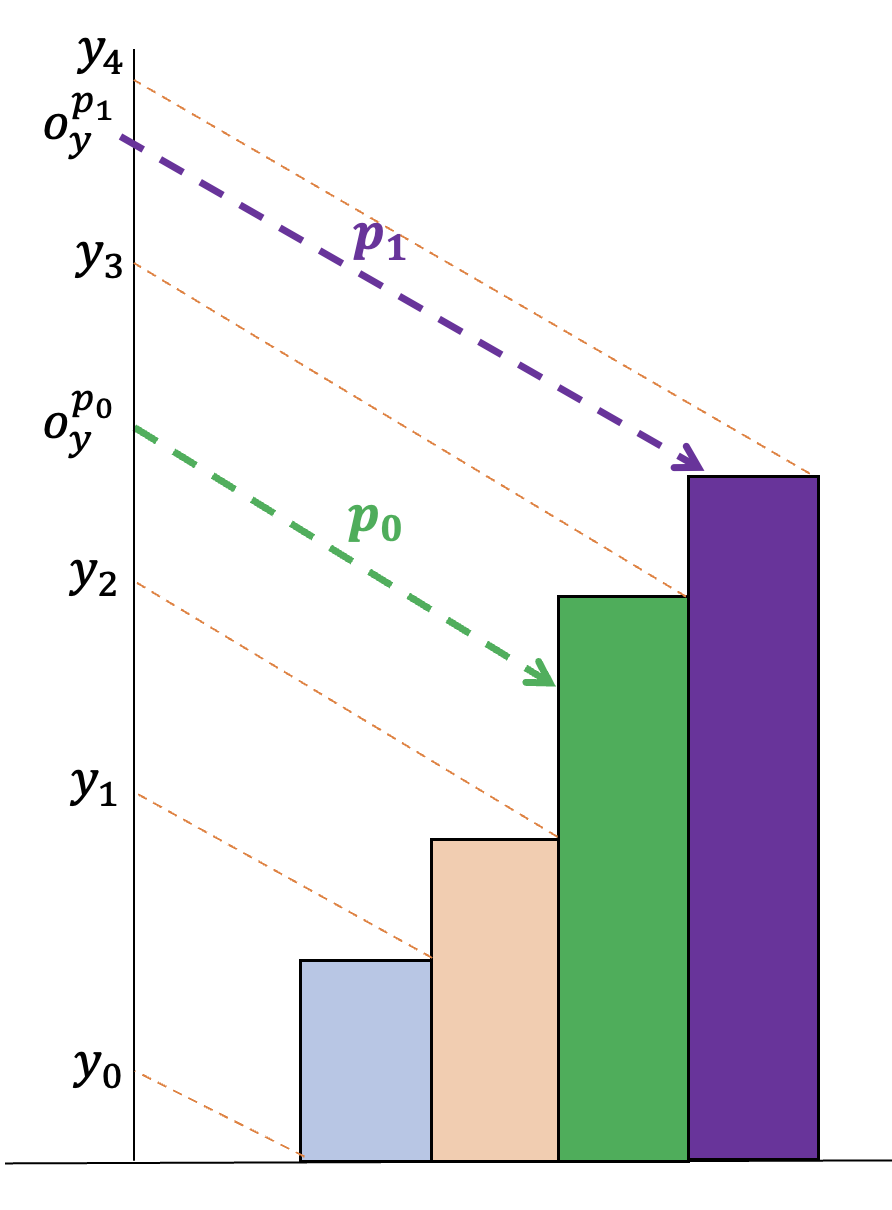}
  \caption{An illustration of two camera rays hitting the surface of the heightfield. Ray $p_0$ will be green, as $y_2 < o_y^{p0} < y_3$. Ray $p_1$ will be purple, as $y_3 < o_y^{p0} < y_4$.} 
  \Description{}
  \label{fig:rays}
\end{figure}

For each camera, we obtain a matrix of camera rays by tracing from evenly distributed points on the $xy$ plane backwards along the direction of the camera. Each camera ray has a direction vector $(d_i, d_j, d_k)$ and origin $(o_i, o_j, o_k)$.

\subsection{Slicing the Heightfield}

To render the resulting image for each camera, we slice the heightfield to obtain a cross-section for each camera ray. We do this by projecting the original 3D ray $\textbf{r}$ onto the $xy$ plane to obtain a 2D projected ray $\textbf{r}_{xy}$. On the $xy$ plane, we determine which heightfield strips this 3D projected vector intersects with and the corresponding coordinates of each intersection.

We then calculate the distance between each adjacent pair of intersection points. We use these distances to construct a slice represented by a 1D array of strip heights $H'$, a cumulative sum of widths $W'$ and colors $C'$. We then re-project the original ray onto this slice to retrieve a final projected ray $\textbf{p}$. The process of projecting a 3D ray to obtain a 2D slice of the heightfield and 2D ray is illustrated in Fig.~\ref{fig:slicing}.

\subsection{Rendering a Single Pixel}

\newcommand{\projray}{p}
\newcommand{\projdirx}{d^\projray_x}
\newcommand{\projdiry}{d^\projray_y}
\newcommand{\projoriginx}{o^\projray_x}
\newcommand{\projoriginy}{o^\projray_y}

We can determine the color of a single ray by determining which strip of the heightfield slice is hit by the projected ray $\textbf{p}$ with direction $(\projdirx, \projdiry)$. We first set the origin of $\textbf{p}$ at $x=0$ and solve for the corresponding $y$ value at this point, to get a projected ray origin $(\projoriginx, \projoriginy)$. 

We then back-trace along the direction of $\projray$ from each strip to find strip boundary parameters $Y = (y_0, \ldots, y_{n+1})$ for each strip, as visualized in Fig.~\ref{fig:rays}. We can calculate $y_0, \ldots, y_{n+1}$ from the sliced strip heights $H_{1D} = h_0, \ldots, h_{n+1}$ and cumulative widths $W_{1D} = (w_0, \ldots, w_{n+1})$ by using the invertible function $f$ defined as
\begin{equation}
y_i = f(i) = h_i - \frac{\projdirx}{\projdiry} \cdot \bigl((i+1)\cdot w_i\bigr).
\end{equation}

\begin{figure}[t]
  \centering
  \includegraphics[trim={0.9in 0in 0.9in 1.5in},clip,width=\hsize]{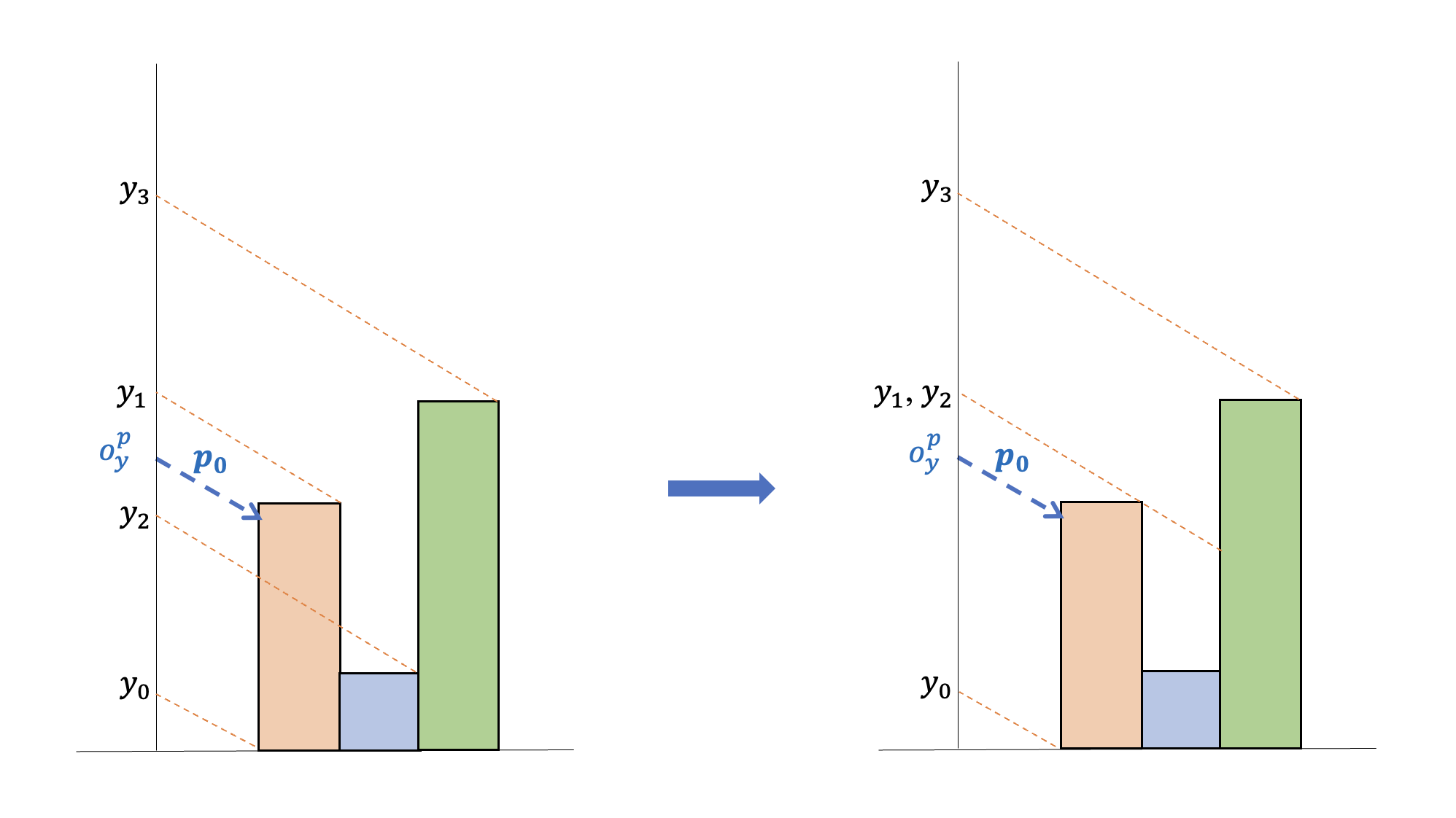}
  \vspace*{-0.2in}
  \caption{Left: An example of non-monotonic back-traced strip heights. 
  Right: The same strips back-traced converted into monotonically increasing heights.}
  \vspace*{0.1in}
  \Description{A woman and a girl in white dresses sit in an open car.}
  \label{fig:monotonic}
\end{figure}

We can determine which strip the ray $\projray$ hits by looking for an $i$ such that, for the pair of boundary parameters $(y_i, y_{i+1})$, we have $y_i < \projoriginy < y_{i+1}$ as shown in Fig.~\ref{fig:rays}. Note that we first need to make sure that the set of boundary parameters is monotonically increasing by using a cumulative maximum, to deal with the case where a strip is entirely obstructed and should be ``skipped'' accordingly, as shown in Fig.~\ref{fig:monotonic}.

We devise a formula for directly obtaining the strip that ray $\projray$ hits based on the reasoning above. Let $g(y)$ be the Heaviside step function where $g(y) = 0$ if $y < 0$ and $g(y) = 1$ if $y \geq 0$. Then, 
\begin{equation}
y_i \leq \projoriginy < y_{i+1} \iff g(\projoriginy - y_{i+1}) == 0,  g(\projoriginy - y_i) == 1
\end{equation}
Thus:
\vspace{-0.1in}
\begin{equation}
y_i \leq \projoriginy < y_{i+1} \iff g(\projoriginy - y_{i}) - g(\projoriginy - y_{i+1}) == 1
\end{equation}
Now, as the $y$  boundary values are monotonically increasing, for the ray $\projray$ with origin height $\projoriginy$ we are guaranteed that we will have $y_i \leq \projoriginy < y_{i+1}$ exactly once across all pairs $(y_i, y_{i+1})$; in all other cases, either $\projoriginy < y_{i}, y_{i+1}$ or $\projoriginy \geq y_{i}, y_{i+1}$, in which case the value $g(\projoriginy - y_{i}) - g(\projoriginy - y_{i+1}) = 0$ (as the values in this subtraction are $0$ or both are $1$). Thus we can calculate the color seen by ray $\projray$ as
\begin{equation}
\begin{aligned}
color(\projray) = &\,\,\ \ \ c_0\bigl (g(\projoriginy - y_0) - g(\projoriginy - y_{1})\bigr) \\
&+c_1\bigl(g(\projoriginy - y_1) - g(\projoriginy - y_{2})\bigr) \\
&+\ \ldots\\
&+ c_n\bigl(g(\projoriginy - y_n) - g(\projoriginy - y_{n+1})\bigr).\
\end{aligned}
\end{equation}
This formula can be rearranged as
\begin{equation}
\begin{aligned}
color(\projray) =&\,\,\ \ \ g(\projoriginy - y_0)\cdot c_0 \\
&+g(\projoriginy - y_1)\cdot (c_1 - c_0)\\
&+\ \ldots\\
&+ g(\projoriginy - y_n)\cdot (c_n - c_{n-1})\\
&-g(\projoriginy-y_{n+1})\cdot c_n.
\end{aligned}
\end{equation}
We thus have a simple formula in terms of backtraced strip heights $y_0, \ldots, y_n$, strip colors $c_0, \ldots, c_n$, and ray height $\projoriginy$ for the view of ray $p$ on a given slice of the heightfield:
\begin{equation}
\begin{aligned}
color(\projray) &=\ \ g(\projoriginy - y_0)\cdot c_0 - H(\projoriginy-y_{n+1})\cdot c_n \\ 
&\ \ +\sum_{i=0}^n g(\projoriginy - y_i)\cdot (c_i - c_{i-1}).
\end{aligned}
\end{equation}

To enable differentiation of the above formula, we use a smooth approximation of the Heaviside step function $g$. More details of this are provided in Section \ref{smooth}.

\subsection{Rendering all the pixels}

For each camera, we render all of the rays coming from the camera with the process above. Note that we can simplify the process somewhat, as for each camera every column of rays will have the same corresponding heightfield slice, so we do not need to re-slice the heightfield for every camera pixel. These rendered rays form an output image $I_i$ for camera $cam_i$.

\subsection{Optimization Process}

Based on the above, the complete process to get from a ray $r_i$ to a view color consists of the following steps:
\begin{enumerate}
    \item Project the ray onto the $xy$ axis and obtain the parameters $H', W', C'$, of the corresponding slice of the heightfield and the new ``2D'' ray $\textbf{p}$.
    \item Apply the differentiable, invertible function $f$ to all of the strips heights $h'_0, \ldots, h'_n$, to obtain backtraced boundary heights $y_0, ... y_1$.
    \item Make the sequence of $y_i$ monotonically increasing, via the function $monotonic(y_i) = \max\{y_0, ..., y_i\}$.
    \item Calculate the color of the ray with $color(p)$.

\end{enumerate}

This establishes a completely differentiable forward rendering process to derive the image seen by a camera pointed at the heightfield. Let this mapping be denoted as
\begin{equation}
I_i = \hbox{\em forward}(cam_i, H, C).
\end{equation}

\paragraph{Repeating Units} There are two general application scenarios of our method: reproducing single images or reproducing repeated patterns. If we wish to reproduce repeated patterns, we optimize for a smaller unit of the heightfield and repeat this unit while considering the occlusion that occurs across units.
\section{Objective Function}

Our objective is to minimize the pixel-wise $MSE$ loss between the actual appearance of the heightfield obtained from our forward rendering algorithm and the given $m$ by $m$ desired appearance images $D_1, \ldots, D_{|CAM|}$:

\begin{equation}
    \min_{H,C} \frac{1}{m^2|CAM|}\sum_{i=1}^{|CAM|}\sum_{x=1}^{m}\sum_{y=1}^{m}\Bigl(Di_{x,y}-\hbox{\em forward}\big(cam_i, H, C\big)_{x,y}\Bigr)^2.
\end{equation}

We also consider several variants on this objective function obtained by adding extra loss terms for surface regularization and different choices for the smooth Heaviside approximation function. We perform an ablation across the regularization variants, which is presented in Table \ref{tab:regularizationablation}. In this ablation, we find that additional loss terms for surface regularization decrease the resulting accuracy of the images on the heightfield's surface. However, this trade-off is necessary to ensure that our surfaces remain within height bounds and do not contain spikes that can easily break or deep troughs that may have extreme shadowing.

\setlength{\tabcolsep}{4pt}

\begin{table}
\small
    \centering
        \caption{MSE loss after 100 steps on two-view optimization}
        \vspace*{-0.05in}
    \begin{tabular}{cccccc}
    \toprule
        \textbf{camera 1} & \textbf{camera 2} & \textbf{none} & \textbf{barrier} & \textbf{smoothing} & \textbf{\hspace*{-0.07in}\begin{tabular}{cc}barrier\\[-0.3ex] smoothing\end{tabular}} \\ \midrule
        black & white & \textbf{0.066} & 0.092 & 0.087 & 0.136\\ 
        black & random & \textbf{0.086} & 0.105 & 0.127 & 0.133 \\ 
        black & stripes & \textbf{0.156} & 0.162 & 0.168 & 0.169 \\ 
        random & stripes & \textbf{0.198} & 0.249 & 0.209 & 0.203 \\
        random & random & \textbf{0.116} & 0.131 & 0.113 & 0.124 \\ \bottomrule
        \vspace{-0.1in}
    \end{tabular}

    \label{tab:regularizationablation}

\end{table}

\subsection{Smooth Heaviside Approximations}\label{smooth}

\begin{figure}
  \centering
  \includegraphics[scale = 0.28]{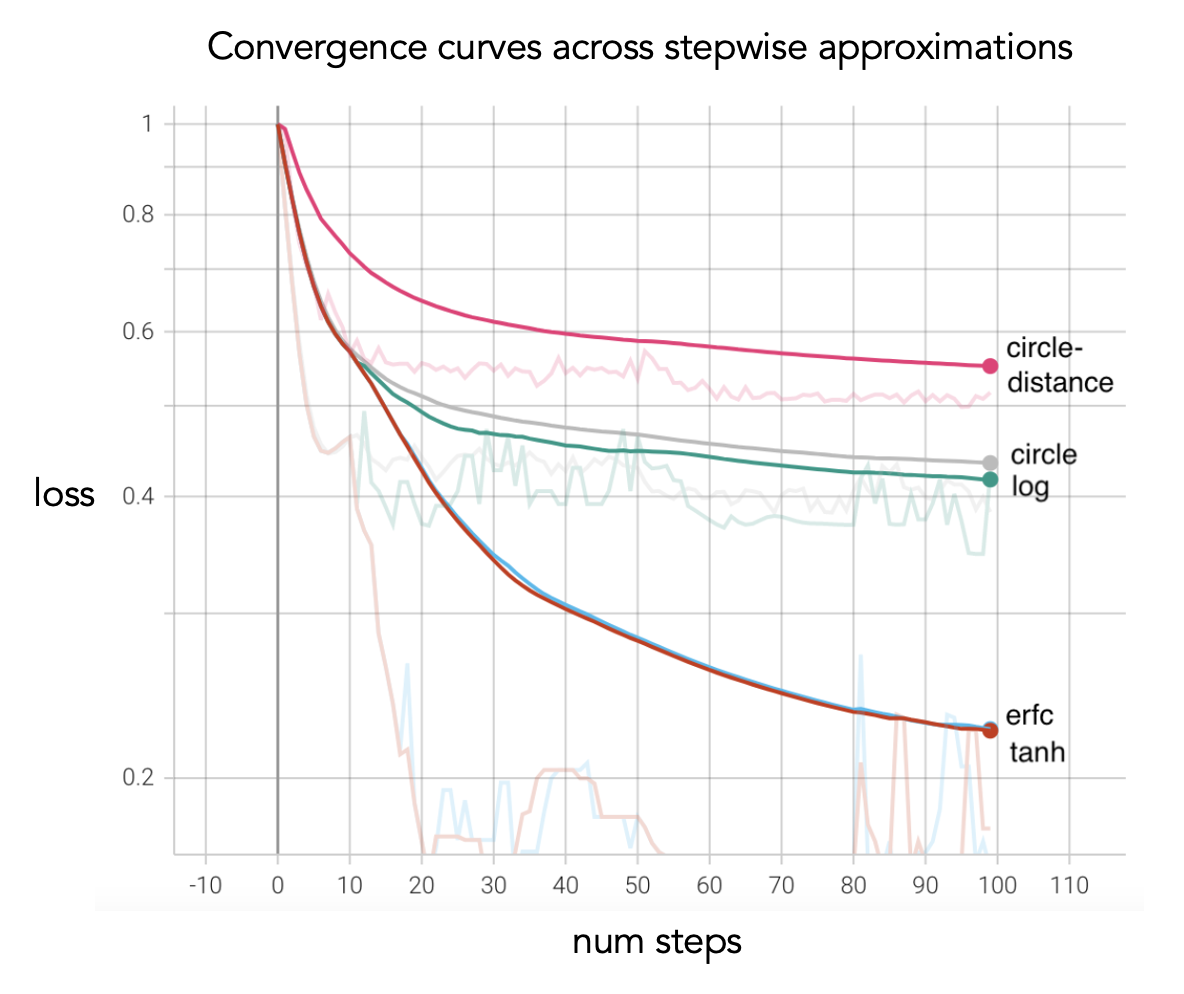}
  \vspace*{-0.1in}
  \caption{Convergence curves for various smooth approximations for the Heaviside stepwise function. erfc and tanh approximations outperform circle distance, circle and log.}
  \label{fig:heavisidecurves}
\end{figure}

The Heaviside step function is not differentiable. We solve this problem by choosing a smooth approximation of this function $\lambda(x,k)$ from a variety of options (outlined in Table \ref{tab:heaviside}) with an additional parameter $k$ which increases the amount of smoothing. We perform an ablation study across various estimation functions with fixed $k = 0.1$, comparing the change in loss over time for a surface optimization over two cameras from opposite sides of the surface with 45 degree elevation. The cameras have desired surface appearance of solid white and solid black respectively. The results of our ablation are presented in Fig.~\ref{fig:heavisidecurves}. Based on this study, we identify the hyperbolic tan ($\hbox{tanh}$) approximation as the most effective approximation for our algorithm.

\begin{table}
\small
  \caption{Heaviside function approximations}
  \vspace*{-0.05in}
  \label{tab:heaviside}
  \begin{tabular}{c@{}c}
    \toprule
    \textbf{Heaviside approximation}\normalsize\strut & \textbf{Formula}\\
    \midrule 
    \\ [-5pt]
    circle &  $\lambda(x,k) = \frac{1}{2} + \frac{1}{2} \
        \frac{x}{(x^2 + k^2)^{1/2}}$\\[5pt]
    circle distance &  $\lambda(x,k) \sim  \hbox{Bern}\left(\frac{1}{2} + \frac{1}{2} \
        \frac{x}{(x^2 + k^2)^{1/2}}\right)$\\[5pt]
    erfc & $\lambda(x,k) = \frac{1}{2} + \frac{1}{2}\left( {1-\frac{2}{\sqrt{\pi}}\int_{0}^{kx} e^{-(kx)^2} \,dx }\right)$\\[5pt]
    tanh & $\lambda(x,k) = \frac{1}{2} + \frac{1}{2}\, \text{tanh}(k x)$\\[5pt]
    log & $\lambda(x,k) \sim \hbox{Bern}\left(\frac{1}{2} + \frac{1}{2}\,\text{tanh}(k x)\right)$\\[5pt]
  \bottomrule
\end{tabular}
\end{table}

\subsection{Regularization}

\paragraph{Barrier Loss}

We regularize our surfaces by enforcing a minimum height $h_{min}$ and maximum height $h_{max}$ via a barrier loss term:
\begin{equation}
\hbox{\em barrier}\_\hbox{\em loss} = -\sum_{i=1}^{|H|}\log(h_{max} - h_i) - \log(h_{min} - h_i).
\end{equation}

\paragraph{Smoothing}

We also add a neighbor loss term for the difference in height between adjacent strips, so that there are no ``spikes'' in the heightfield:
\begin{equation}
neighbor\_loss = -\sum_{i=1}^{|H|} |h_i - h_{i-1}|. 
\end{equation}

\begin{table}[h]
\small
    \centering
    \caption{Loss after 100 steps on surfaces with uniform initial configurations, for different combinations of optimizations.}
    \vspace*{-0.05in}
    \begin{tabularx}{\columnwidth}{cccccc}
    \toprule
        \textbf{cam 1} & \textbf{cam 2} & \textbf{none~~} & \textbf{\begin{tabular}{@{}c@{}}coarse \\[-0.3ex] to fine\end{tabular}} & \textbf{\begin{tabular}{@{}c@{}}coord. \\[-0.3ex] descent\end{tabular}} & \textbf{\begin{tabular}{@{}c@{}}coarse to fine\\[-0.3ex] coord.~descent\\[-0.3ex] sim.~annealing\end{tabular}} \\ \midrule
        black & white & 0.136 & 0.067 & 0.107 & \textbf{0.011} \\ 
        random & random & 0.124 & 0.099 & 0.108 & \textbf{0.093} \\ 
        black & random & 0.133 & 0.108 & 0.094 & \textbf{0.051} \\ 
        black & stripes & 0.169 & 0.124 & 0.141 & \textbf{0.037} \\ 
        random & stripes & 0.203 & 0.205 & 0.180 & \textbf{0.057} \\ \bottomrule
    \end{tabularx}
    \vspace*{0.05in}
    \label{tab:optablation}
\end{table}

\vspace{-0.2in}
\section{Optimization Methods}

Our basic optimization algorithm uses the Adam Optimizer \cite{Kingma2015} to minimize the MSE loss between the desired views and actual appearance of the surface for each camera. To design our surfaces, we introduce a suite of various optimization techniques. We perform an ablation across these methods, which is presented in Table \ref{tab:optablation}. We find that a combination of coarse-to-fine optimization, alternating block coordinate descent, and simulated annealing significantly improves performance in all test cases.

\subsection{Coarse-To-Fine Optimization}

We use coarse-to-fine optimization, whereby every 50 steps of the algorithm, each of the heightfield strips is subdivided into a $2 \times 2$ block of strips with the same height and color as the original strips. We begin the optimization process with heightfields of $8 \times 8$ strips, and end the process with heightfields of $32 \times 32$ strips. This provides some additional surface regularization and also improves the overall performance of the algorithm.

\subsection{Alternating Block Coordinate Descent}

During optimization, we alternately update heights of the strips for 10 steps and the colors of the strips for 20 steps, until convergence. This greatly improves the performance of our algorithm relative to optimizing both heights and colors simultaneously at each iteration.

\subsection{Simulated Annealing}
To improve our results for more complicated cases and better escape local optima, we optionally perform steps of simulated annealing \cite{bertsimas1993simulated}, as outlined in Algorithm \ref{algo:simannealing}, on both the strip heights and colors at the start of the algorithm and subsequently every 100 steps. We set $T_{max} = 3$, $T_{min} = 0.5$ and cool $T$ using exponential decay with a factor of $0.99$. If a more uniform final pattern is desired, then simulated annealing can be omitted from the optimization process.

\begin{algorithm}
\small
\caption{\strut~\small Simulated Annealing~~\footnotesize\cite{bertsimas1993simulated}}\label{algo}
\begin{algorithmic}
\vspace{0.5ex}
\State $T\gets T_{max}$
\State $C, H\gets$ init\_C, init\_H{$ $}
\vspace{1ex}
\While{$T>T_{min}$}
\vspace{1ex}
	\State $C_{rand}, H_{rand}\gets $ \textbf{\Call{RANDOM\_NEIGHBOUR}{$T, C, H$}}
    \vspace{0.3ex}
	\State $\Delta L\gets$ \textbf{\Call{LOSS}{$C_{rand}, H_{rand}$}} $-$ \textbf{\Call{ENERGY}{$C_{rand}, H_{rand}$}}
	\vspace{1ex}
    \If{$\Delta L < 0$}
      \State $COL, H\gets C_{rand}, H_{rand}$
	\vspace{1ex}
    \ElsIf{\Call{random}{$ $} $<$ $e^{-\Delta L/T}$}
		\State $C, H\gets C_{rand}, H_{rand}$
	\EndIf
	\vspace{1ex}
    \State $T\gets$ \textbf{\Call{COOL}{$T, C, H$}}
\EndWhile
\\ \vspace{1ex}
\Return $C, H$
\end{algorithmic}
\label{algo:simannealing}
\end{algorithm}

\begin{figure}
  \centering
  \vspace*{0.25in}
  \includegraphics[scale = 0.3]{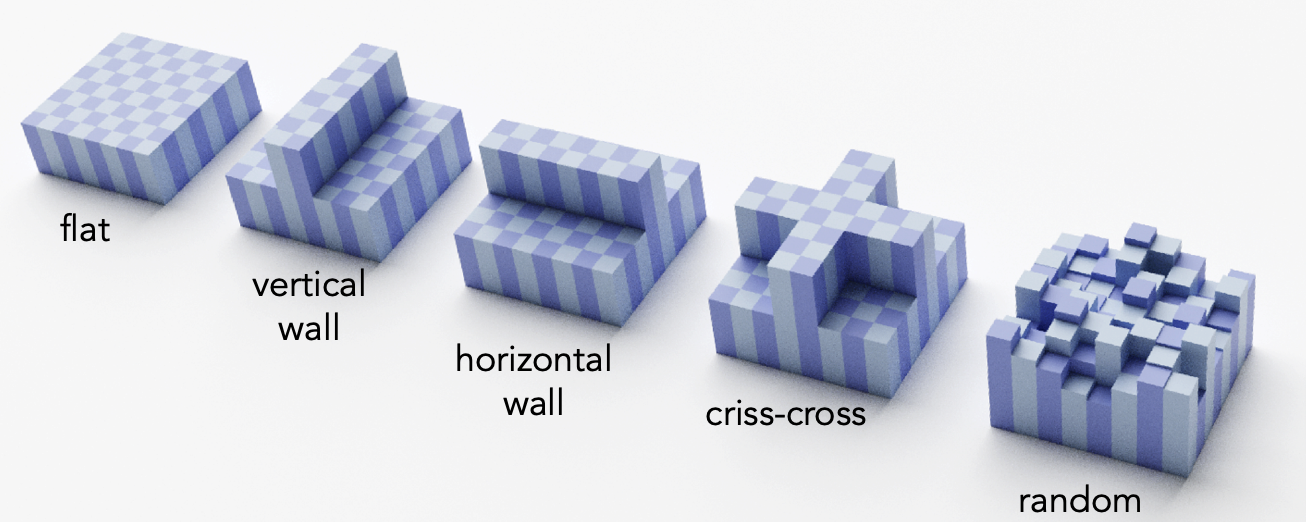}
  \caption{Renders of $8 \times 8$ initial surface configurations (left to right: flat, vertical wall, horizontal wall, cross, random).}
  \Description{}
  \label{fig:init}
\end{figure}

\begin{figure}
  \centering
  \includegraphics[scale=0.3]{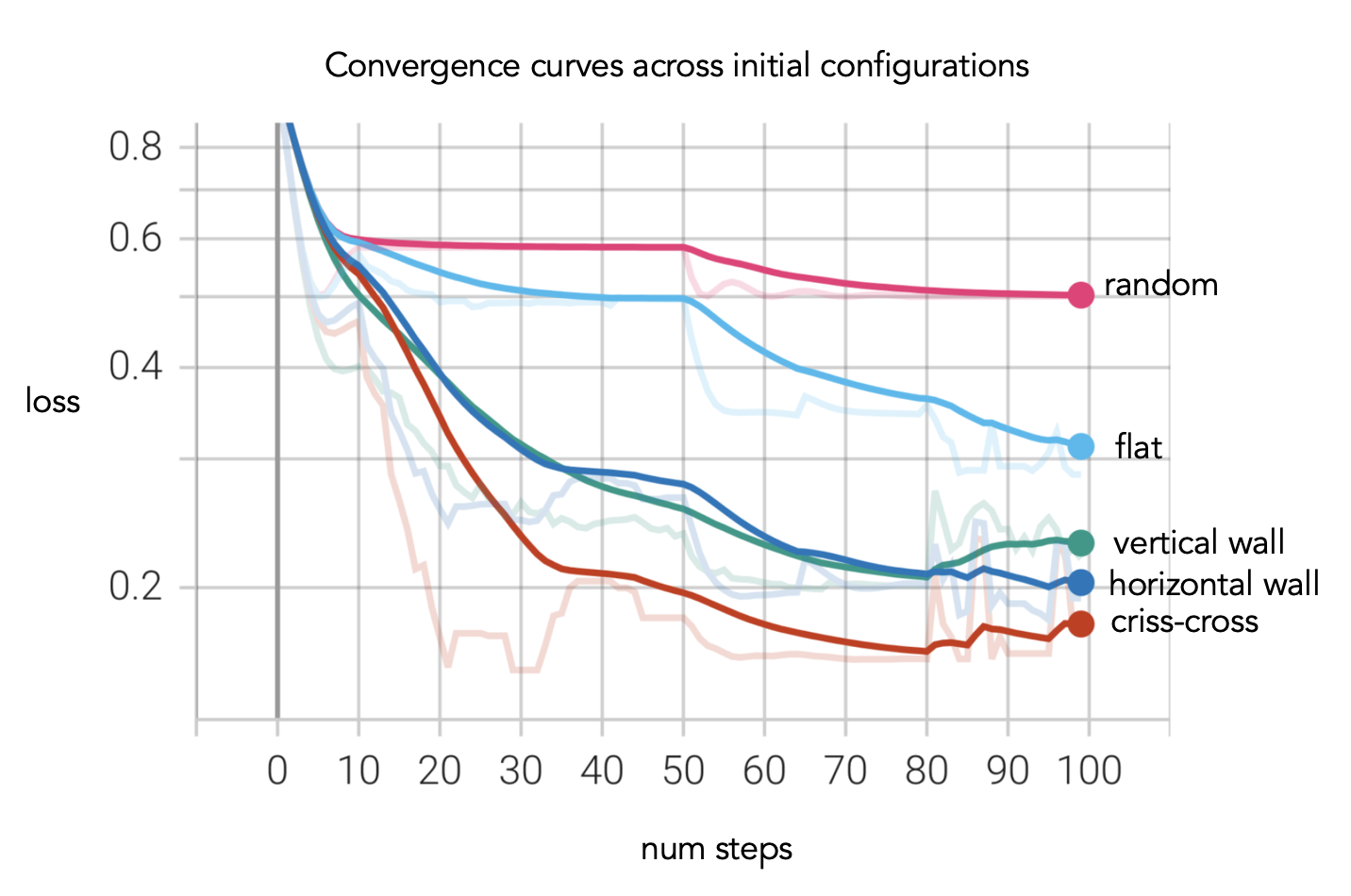}
  \vspace*{-0.22in}
  \caption{Comparison of convergence curves across different initial surface configurations.}
  \Description{}
  \label{fig:initstudy}
\end{figure}

\subsection{Initial Configurations}

We experiment with several different initial configurations for our optimization, including uniform, random, and preset configurations, as illustrated by Fig.~\ref{fig:init}. We compare the convergence curves when using these initial configurations for the four-view optimization problem (with elevation 30 and desired appearances of solid cyan, magenta, yellow and black). The results of this experiment are shown in Fig.~\ref{fig:initstudy}. We find that using a criss-cross initial configuration gives the best performance, closely followed by vertical and horizontal wall configurations (which, aside from randomness, should theoretically all be equivalent for this four-view problem).

\subsection{Post-optimization projection}

As an additional step after optimizing the heights and colors of the field, we create multiple vertical segments for each heightfield bar, and then re-project the desired images onto the heightfield. This allows us to achieve higher resolution results without additional optimization, which is particularly important for detailed images.

\begin{figure}
  \centering
  \includegraphics[width=\hsize]{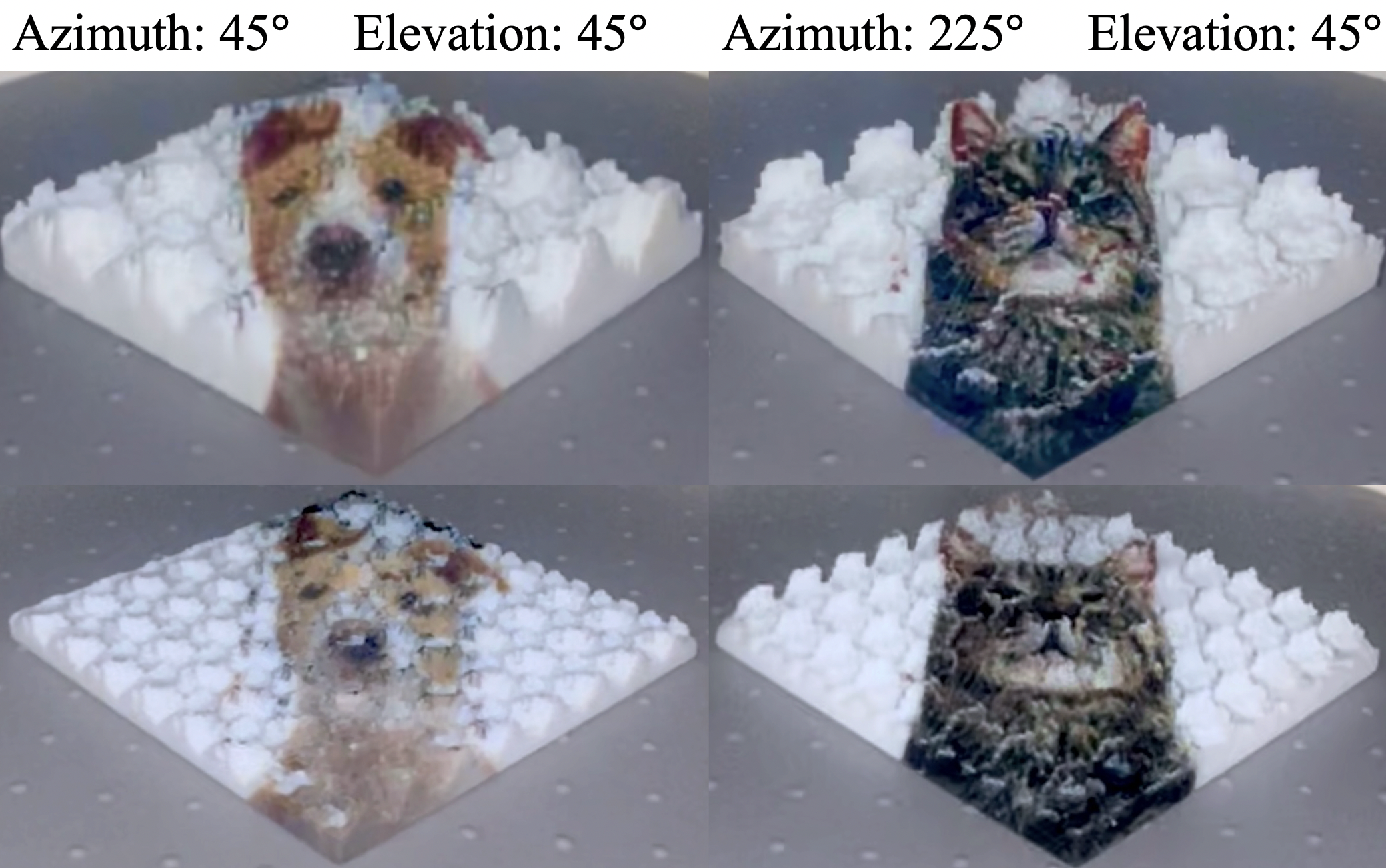}
  \caption{Printed surfaces viewed from two directions, each displaying two images.  The surface printed with lower bar width (bottom) suffers from more color mixing due to translucency, compared to the one with higher bar width (top).}
  \Description{}
  \label{fig:dogcat}
\end{figure}

\section{Optimization Results}

We implemented the optimization algorithm using {\em PyTorch}, and the fabrication mesh conversion script in Blender with the Python {\em bpy} module. Optimization times range from 30 minutes to 2 hours.

\begin{figure*}[h]
  \centering
  \vspace*{0.05in}
  \includegraphics[scale = 0.8]{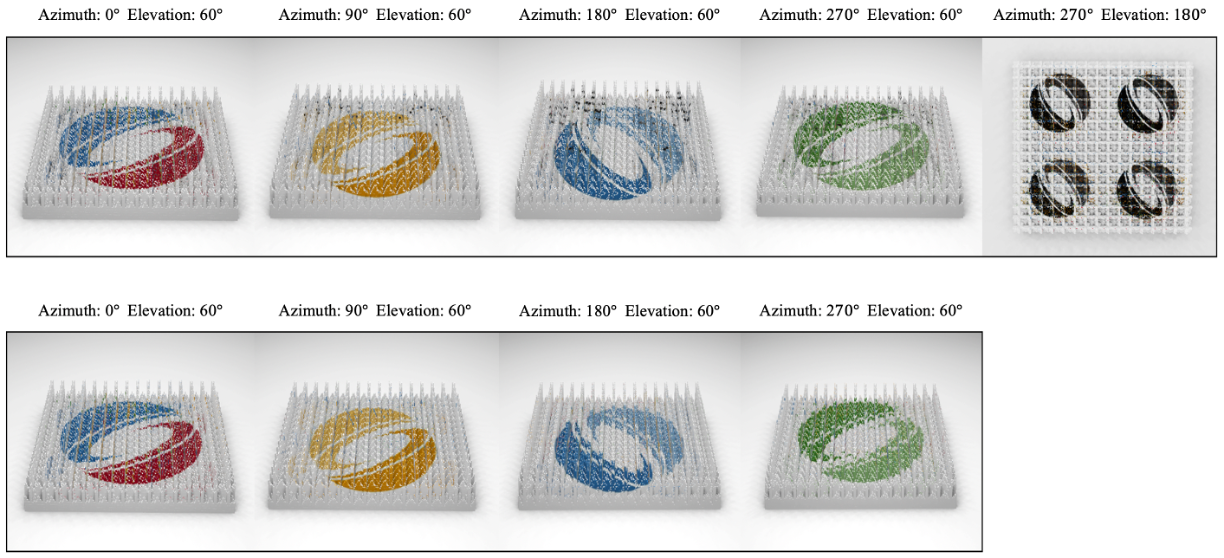}
  \caption{Top: Optimized surface with five distinct desired appearances. Bottom: Optimized surface with four distinct desired views.}
  \Description{}
  \label{fig:siggraph}
  \vspace*{0.1in}
\end{figure*}

We perform optimizations for various test cases, with varying numbers of cameras, viewing angles and desired appearance. In Fig.~\ref{fig:siggraph}, we show surfaces optimized for five views (top) and four views (bottom). The results closely match the desired appearance, but we can see some quality degradation occurring when introducing the fifth view. We show additional rendered heightfields in our video.

\begin{figure*}[h]
    \captionsetup[subfigure]{labelformat=empty}

    \centering
    \begin{subfigure}[t]{0.45\textwidth}

        \centering
        \includegraphics[width=\linewidth]{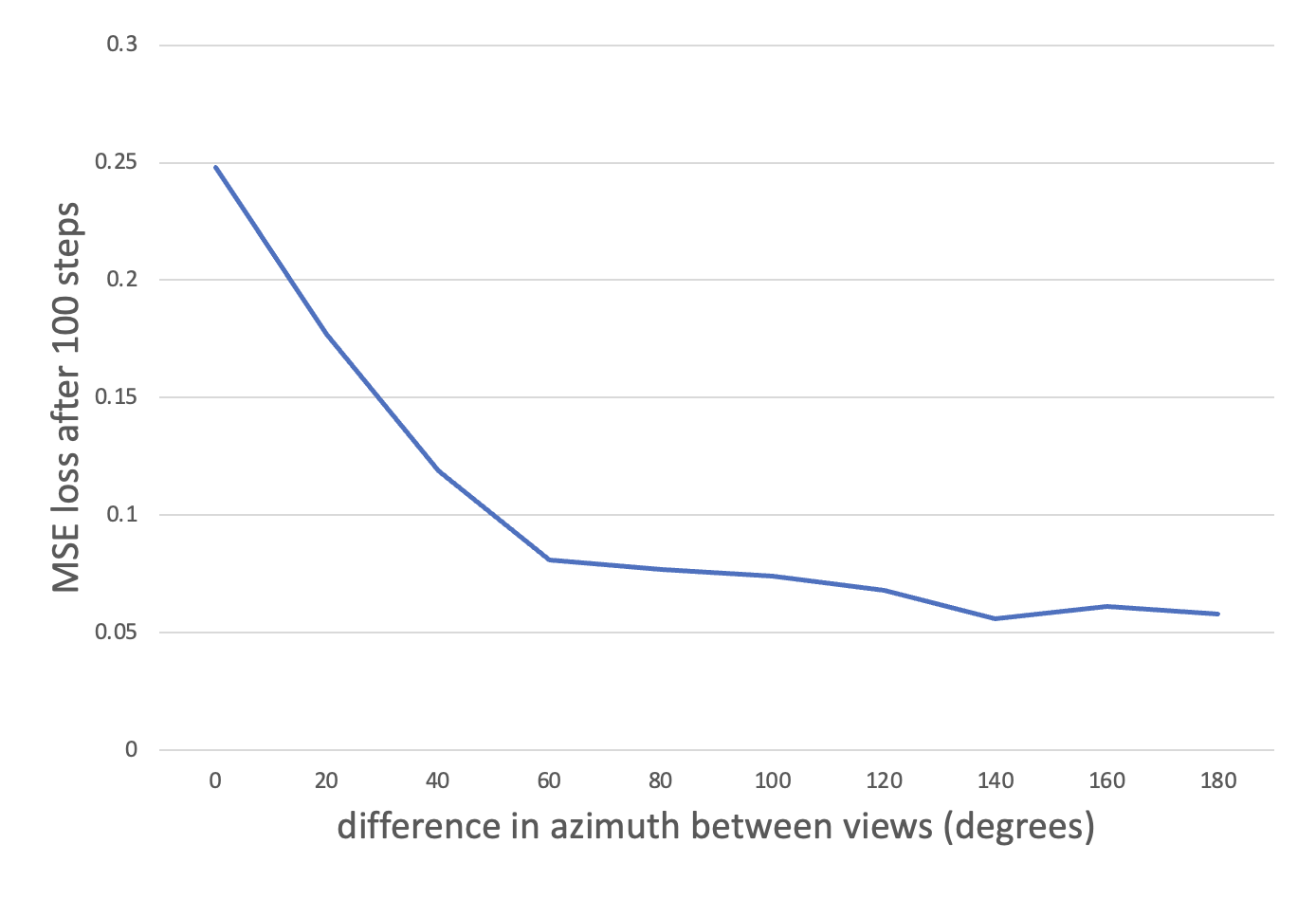} 
        \caption{Fig. 13a. MSE loss after 100 steps with two viewing directions. The first viewing direction has a desired appearance of solid black, with elevation 60 and azimuth 0. The second viewing direction has a desired appearance of solid white, elevation 60 degrees and variable azimuth.} \label{fig:azimuth}
    \end{subfigure}
    \hfill
    \begin{subfigure}[t]{0.45\textwidth}

        \centering
        \includegraphics[width=\linewidth]{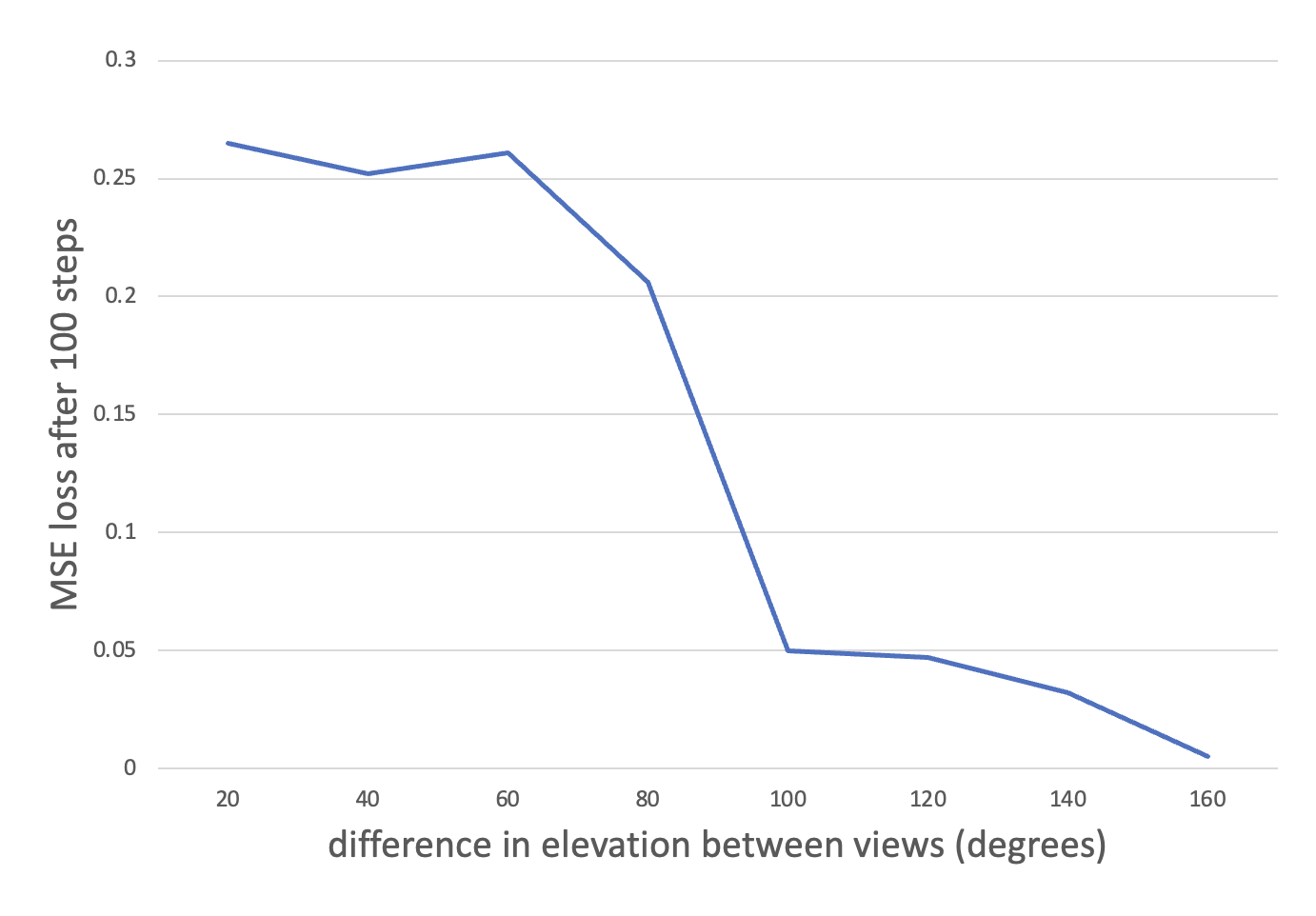} 
        \caption{Fig. 13b. MSE loss after 100 steps with two viewing directions. The first viewing direction has a desired appearance of solid black, with elevation 20 and azimuth 0. The second viewing direction has a desired appearance of solid white, variable elevation and azimuth 0.} \label{fig:elevation}
    \end{subfigure}

\end{figure*}

We also carry out a study to determine which pairs of viewing angles are compatible with each other. We examine the loss after 100 steps on a two-view (solid black vs. solid white) optimization. In Fig.~\ref{fig:azimuth} both views have the same elevation and different azimuth, and in Fig.~\ref{fig:elevation} both views have the same azimuth and different elevation. We find that as difference in elevation and azimuth increases beyond 40 degrees and 60 degrees respectively, we are able to achieve a final MSE loss of $<0.1$.

\begin{figure*}[h]
  \centering
  \vspace*{0.05in}
  \includegraphics[width=\linewidth]{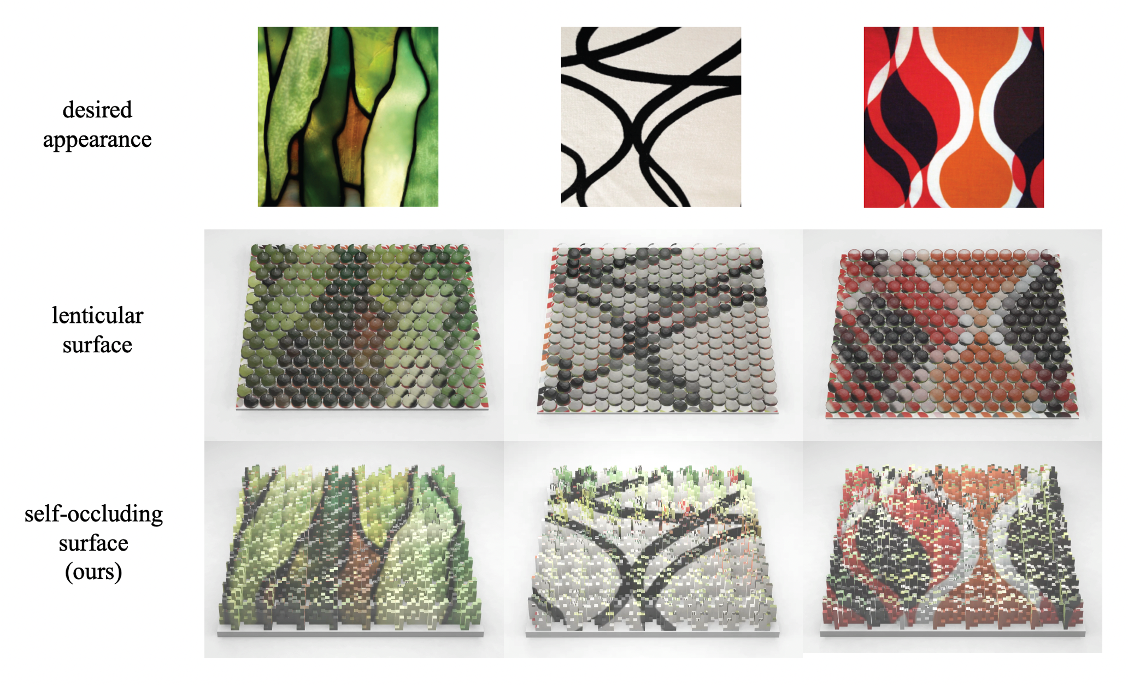}
  \caption{Top: desired appearances. Middle: Renders of lenticular surface from three different viewing directions. Bottom: Render of self-occluding surface generated via our method from three different viewing directions.}
  \Description{}
  \label{fig:lens}
\end{figure*}

Additionally, we compare our approach to a lenticular-based approach~\citep{zeng2021lenticular}, and present renders of resulting surfaces in Fig.~\ref{fig:lens} that reflect the highest possible resolution for each approach with the same 3D printer. We find that our approach is qualitatively better at capturing sharp lines and high-resolution details from the desired appearance images, although it does suffer slightly more from color contamination across views. We also note that our method displays a smaller crop of the desired appearance image than the lenticular surface, due to the need for cropping during image projection onto the heightfield surfaces.

\section{Fabrication}

\begin{figure*}[h]
  \centering
  \vspace*{0.05in}
  \includegraphics[width=\linewidth]{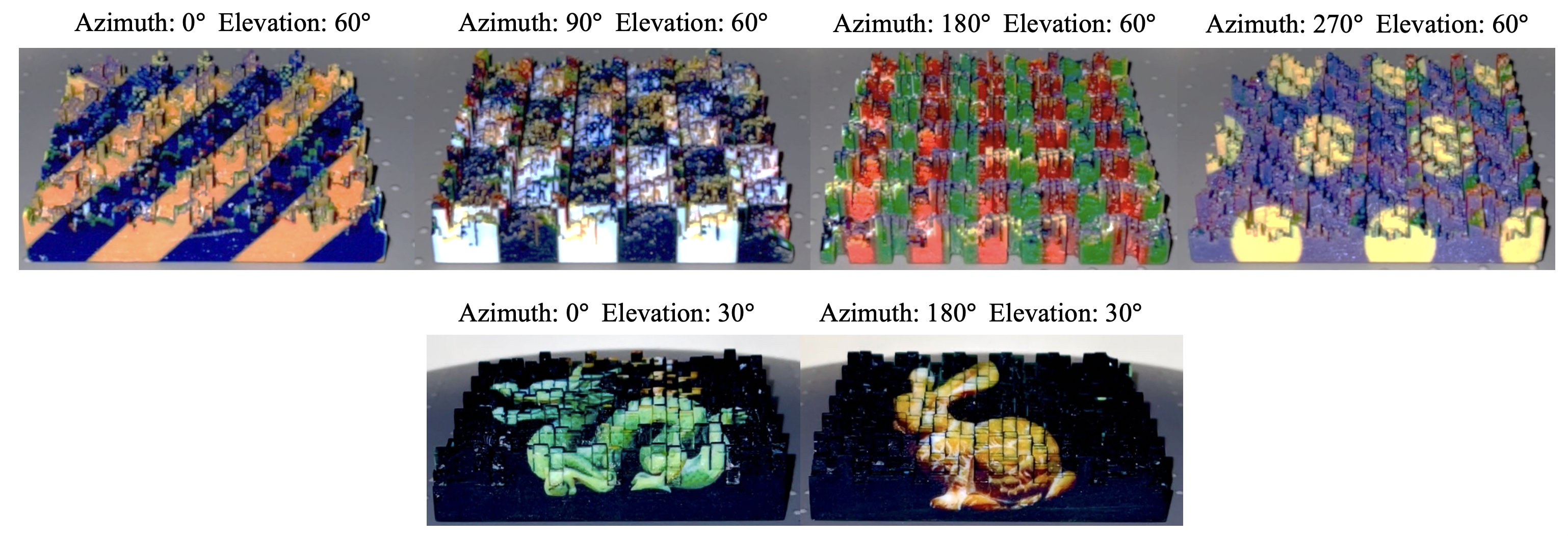}
  \caption{Photographs of various fabricated surfaces. Top row: The same surface as viewed from four directions, displaying four unrelated colorful patterns. Bottom row: The same surface as viewed from two directions, displaying the Stanford Dragon and the Stanford Bunny.}  
  \Description{}
  \label{fig:phys}
  \vspace*{0.1in}
\end{figure*}

\begin{figure*}[h]
    \captionsetup[subfigure]{labelformat=empty}
    \centering
    \begin{subfigure}[t]{0.45\textwidth}
        \centering
        \includegraphics[scale = 0.2]{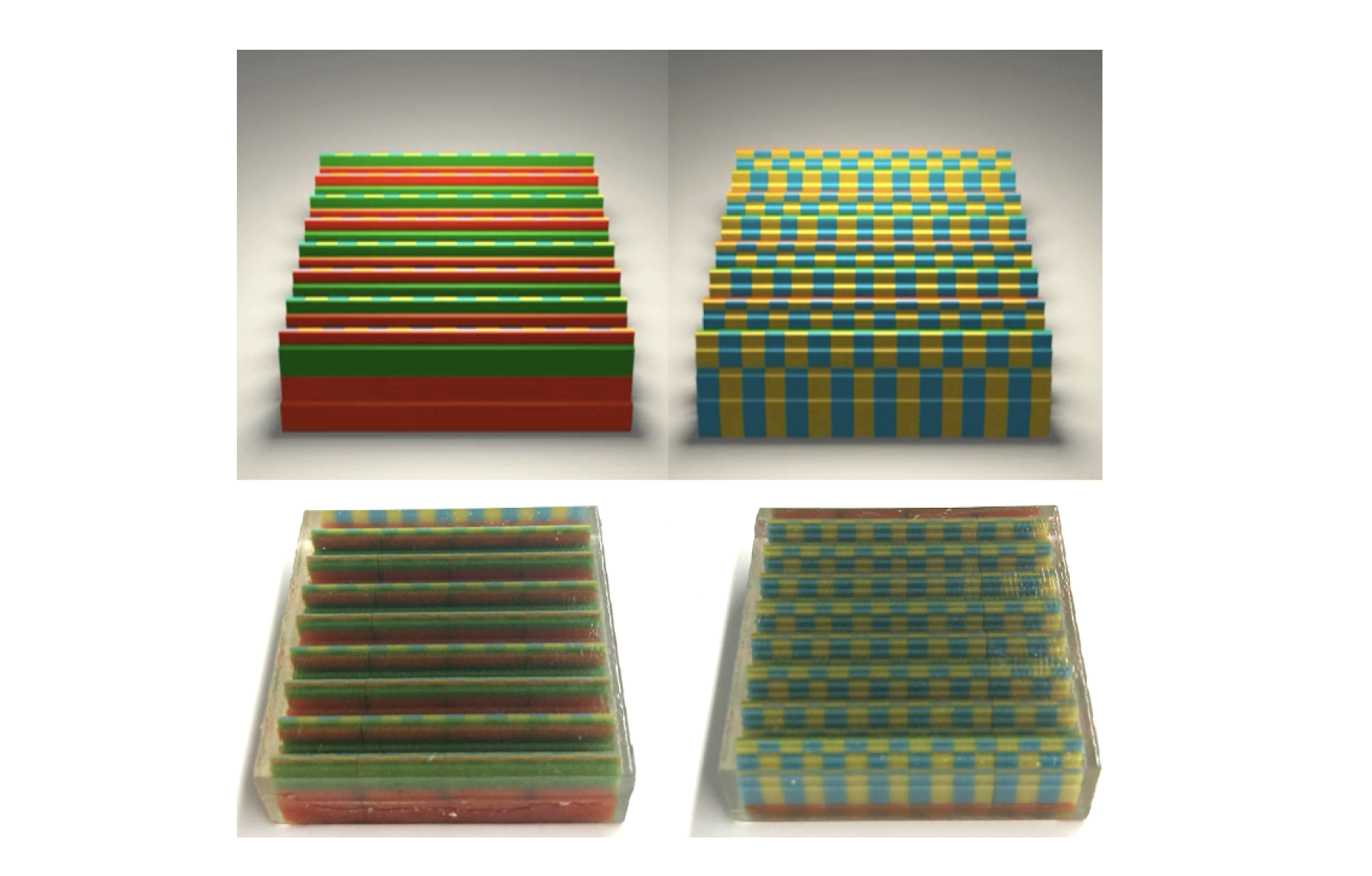} 
        \caption{Fig. 14a. Top: Render of synthetic surface as viewed from two directions. Bottom: Fabricated version of this same surface.} \label{fig:simtoreal}
    \end{subfigure}
    \hfill
    \begin{subfigure}[t]{0.45\textwidth}
        \centering
        \includegraphics[width=\linewidth]{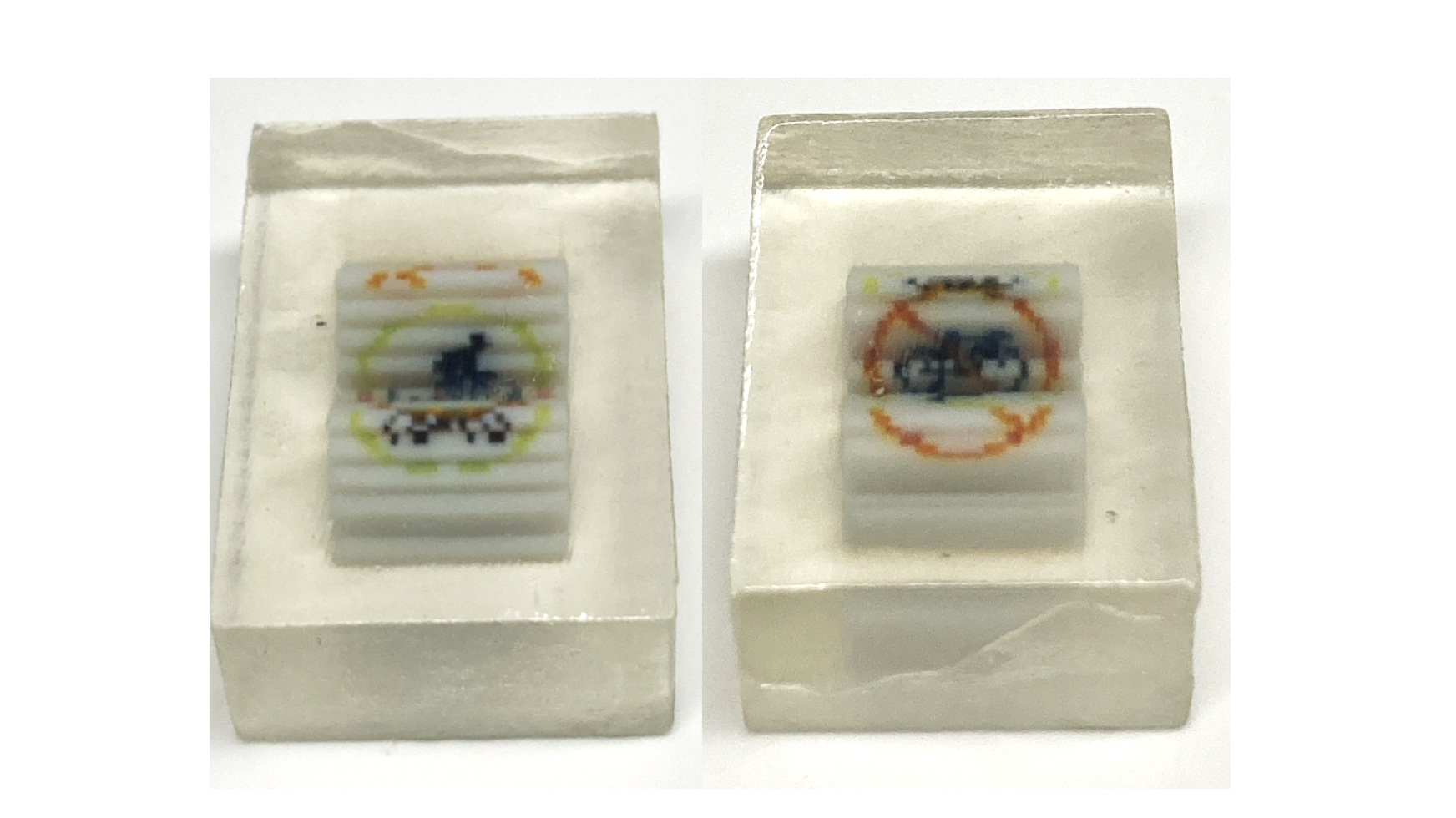} 
        \caption{Fig. 14b. High-resolution fabrication of a biking/no biking sign as viewed from two high-elevation directions  (60 degrees). The surface is less than 1cm wide.} \label{fig:bike}
    \end{subfigure}
\end{figure*}

To fabricate the surfaces, we first convert the heightfields into meshes and material template libraries using Blender. We then fabricate the surfaces using a Stratasys J55 Polyjet UV printer. Results are shown in Fig.~\ref{fig:dogcat} and Fig.~\ref{fig:phys}. An example of a rendered and fabricated version of the same surface can be seen in Fig.~\ref{fig:simtoreal}, which also shows a close match between rendered and real results. We identified the best workable resolution for the printer by performing several strip tests, and found that 300dpi was possible without aliasing. We used this resolution as the minimum strip width for our fabrication.
We are able to achieve view-dependent effects at a very high resolution as can be seen in Fig.~\ref{fig:bike}, which is less than 1cm wide. In order to prevent strip breakage for such  high-resolution prints, we added a layer of ultra-clear printed material on top of this surface. This layer, however, reduces the actual elevation required for the desired surface appearance due to refraction.

Some results also suffer from color mixing across views due to undesired printing material translucency, such as in Fig.~\ref{fig:dogcat}. Our current approach to solve this is to reduce the printing resolution for a given target. Future work may also regularize the heightfield by minimizing high color variance between neighboring bars to further mitigate the negative effects of material translucency without having to compromise on printing resolution.

\section{Discussion and Limitations}

Our algorithm explores the use of self-occluding heightfields in fabricating multi-image displays. We demonstrate generating surfaces with view-dependent appearance at up to five distinct viewing angles and fabricate surfaces that closely match the rendered results at high resolution and with up to four viewing angles. The most important advantages of our method as opposed to prior works are the high working resolution and ability to share colors across views. We are also able to use bright colors on our surfaces, as we do not rely on additional walls that darken the overall surface appearance as in \citet{sakurai2018fabricating}. However, we observe that the quality of results from our procedure is highly dependent on the relative azimuth and elevation of the cameras, unlike with lenticular-based methods. Additionally, we do not take the self-shadowing of environment lighting into account. Future work could, however, augment our algorithm to account for self-shadows by darkening or lightening the colors in the heightfield to offset these effects.

\section{Conclusion}

We devise a novel approach for fabricating multi-image displays that does not rely on building fixed-color walls or using lenses and polish. We present a suite of techniques that comprise a new optimization algorithm specifically tailored to our task, including a simple differentiable ray-casting inspired rendering algorithm designed to render colored heightfields to achieve our task. Our approach allows us to use a UV printer to successfully fabricate colorful 3D objects whose surface-appearance changes depending on viewing angle. 

\begin{acks}
We thank Baffour Osei, the Princeton Robotics Lab, and Princeton SEAS for their help and support during this project.
\end{acks}

\bibliographystyle{ACM-Reference-Format}
\bibliography{sample-base}
\newpage

\end{document}